\documentclass[twocolumn,showpacs,preprintnumbers,amsmath,amssymb,prappl,superscriptaddress]{revtex4-1}
\usepackage{graphicx}
\usepackage{dcolumn}
\usepackage{mathrsfs}
\usepackage{bm}
\usepackage{color}

\begin{document}

\preprint{}

\title{Light Sheets with Extended Length}

\author{Shengwang Du}\email{Corresponding author: dusw@ust.hk}
\affiliation{Department of Physics, The Hong Kong University of Science and Technology, Clear Water Bay, Kowloon, Hong Kong, China}

\author{Teng Zhao}
\affiliation{Light Innovation Technology Ltd, Hong Kong, China}

\author{Luwei Zhao}
\affiliation{Light Innovation Technology Ltd, Hong Kong, China}

\date{\today}

\begin{abstract}
High resolution light-sheet optical microscopy requires illuminating a sample by a thin excitation light sheet with large area.  Here, we describe methods for producing light sheets with extended length, as compared to Gaussian-beam-type diffracting light sheets with the same thickness, without scanning or dithering the light beams. 
\end{abstract}

\maketitle

\section{Introduction}\label{sec:Introduction}

In a conventional wide-field fluorescence optical microscope, the excitation and imaging optics share the same objective lens and the excitation light passes through the specimen. The out-of-focus fluorescence thus adds background noise to the image. Confocal microscopy solves this problem by focusing the excitation light into a spot and rejecting the out-of-focus fluorescence light by a pinhole in the detection side. By scanning the focus point by point, the confocal microscopy allows for reconstructing the object's three-dimensional (3D) structure. In such a scanning confocal microscope, although 3D high-resolution information can be obtained, illumination beam goes through the entire thickness of the specimen, which causes serious phototoxicity to the sample and  photobleaching to the fluorophore. Besides, confocal microscopy is slow because of its point-by-point scanning. Therefore, confocal microscopy by its nature is not suitable for 3D live-cell imaging though many efforts have been made to improve its performance \cite{Stephens2003}. 

Recent development of light-sheet microscopy (LSM) has demonstrated its capacity for live imaging of biological samples from single cells to tissues with high spatial and temporal resolutions \cite{OlarteTutorial2018, Stelzer2015}. By illuminating the excitation light sheet from side perpendicular to the imaging objective axis, a light-sheet fluorescence optical microscope has the optical sectioning capability with ultra low phototoxicity. It is also much faster than a conventional confocal microscope because LSM unitizes plane-scanning rather than point-scanning. LSM is a powerful tool for four-dimensional (4D: 3D in space plus 1D in time) live-sample imaging.

In fluorescence LSM, the excitation light sheets is expected to be thin in axial dimension along the optical axis of detection objective (to eliminate off-focus excitation), and also be long in lateral dimension such that a lager field of view can be covered. However, these two conditions contradict each other for the conventional LSM based on Gaussian beams because of the diffraction effect: a thinner Gaussian beam propagates a shorter distance. For a two-dimensional (2D) Gaussian-beam light sheet with a thickness approaching the diffraction limit of half wavelength, the propagation length becomes shorter than a few wavelengths, in which the light sheet is too short to even cover a single cell. Non-diffracting Bessel beams has been utilized in order to produce longer light sheets. In these cases Bessel beams are scanned into a time-averaged sheet. While the scanning Bessel-beam light sheet indeed extends the propagation length significantly, its strong side lobes induce strong off-focus excitation and increases photo-toxicity\cite{SBS}: the Bessel beam has its energy evenly distributed among its rings \cite{DurninOL1998, McGloin2005}. In reality, the propagation length is limited by the inverse of the finite angular spectrum width. 

If the light energy of a Bessel beam is spread into multiple beams on the sheet, such a light sheet has much lower instantaneous intensity than a scanning Bessel-beam light sheet under the same illumination laser power, and thus leads to much lower nonlinear photodamage and phototoxicity to the specimen. This effect was demonstrated by Gao \textit{et. al.}  using a linear array of Bessel beams \cite{GaoCell2012}. It was later extended to 2D optical lattice light sheet microscopy (LLSM) for achieving higher resolution and lower phototoxicity \cite{LatticeLightSheet}. However, LLSM is complicated to implement, has too many degrees of freedoms for optimization, and has limited access only within very a few groups. To produce a time-averaged uniform light sheet, LLSM requires dithering its 2D lattice pattern.

The phototoxicity can be further reduced by spreading the light energy uniformly into the sheet. We recently demonstrated a simple and robust method to generate ultrathin line Bessel sheets (LBS) \cite{LBS, Du2018}. Using this technique, we create smooth and long light sheets without dithering the beam, which can have much less phototoxicity than structured light sheets. In this article, we reveal the theory and design principle of the LBS, as well as for producing more general 2D smooth light sheets with extended length with two-beam interference. Here we focus only the spatially-smooth light sheets without need of scanning or dithering the beams.

The article is organized as follows. In Sec.~\ref{sec:GeneralFormalism}, we provide the general formalism of 2D light sheets. In Sec.~\ref{sec:DLS}, we describe the properties of diffracting light sheets. Then we derive and describe non-diffracting light sheets with extended length by two-beam interference in  Sec.~\ref{sec:NDLS}, where we also show the numerical comparison between the diffracting and non-diffracting light sheets. We conclude in Sec.~\ref{sec:Summary}.

\section{General Formalism of 2D Light Sheets} \label{sec:GeneralFormalism}

We choose the coordinate system in which the coherent laser light, with wavelength $\lambda$, propagates along $z$ direction. The light sheet is confined in $x$ direction with a thickness $d$ and spread over $y$ direction with a width $t$. Ideally, to achieve an absolute non-diffracting beam, its angular spectrum in $k_x-k_y$ plane must be confined on a circular ring, \textit{i.e.}, $k_z=\sqrt{k_0^2-k_x^2-k_y^2}$ is a constant where $k_0=2\pi/\lambda$, so that the beam profile does not change along its propagation. The zeroth-order Bessel beam is a specific case with even distribution on the ring. For a real case, the circular ring always has a finite spectrum width that leads to a finite propagation length. In this article, we work under the wide sheet condition $t\gg\{d, \lambda\}$ such that the light (complex) electric field can be described as $f(x,z)$ and its angular spectrum is given as $F(k_x)\delta(k_y)$. From now on, we will focus on the dimensions $x$ and $z$ only.

In Fourier optics, the scalar complex electric field $f(x,z)$ and its angular spectrum $F(k_x)$ is connected by the following optical Fourier transform
\begin{equation}
f(x,z)=\frac{1}{\sqrt{2\pi}}\int_{-k_0}^{k_0}F(k_x)e^{ik_xx}e^{i\sqrt{k_0^2-k_x^2}z}dk_x,
\label{eq:GeneralEq}
\end{equation}
where $k_x$ is confined within $[-k_0,k_0]$ for a fixed wavelength $\lambda$. The angular spectrum can be obtained from the following inverse Fourier transform
\begin{equation}
F(k_x)=\frac{1}{\sqrt{2\pi}}\int_{-\infty}^{\infty}f(x,0)e^{-ik_xx}dx.
\label{eq:InverseFourier}
\end{equation}
For mathematic convenience, we take the following normalization (from energy conservation) for the discussion in this article  
\begin{equation}
\int_{-\infty}^{\infty}|f(x,z)|^2dx=\int_{-k_0}^{k_0}|F(k_x)|^2dk_x=1.
\label{eq:Nomalization}
\end{equation}

When the wave field envelope varies slowly as compared to its wavelength, \textit{i.e.}, $|k_x|\ll k_0$, we can take the paraxial approximation
\begin{equation}
\sqrt{k_0^2-k_x^2}z\simeq k_0z-\frac{k_x^2}{2 k_0}z,
\label{eq:ParaxialApproximation}
\end{equation}
and rewrite Eq. (\ref{eq:GeneralEq}) as
\begin{equation}
f(x,z)=\frac{1}{\sqrt{2\pi}}\int_{-k_0}^{k_0}\tilde{F}(k_x,z)e^{ik_xx}dk_x,
\label{eq:GeneralEq2}
\end{equation}
where the z-dependant spectrum is
\begin{equation}
\tilde{F}(k_x,z)=F(k_x) e^{i\sqrt{k_0^2-k_x^2}z}\simeq F(k_x) e^{ik_0z} e^{-i\frac{k_x^2}{2k_0}z}.
\label{eq:F2}
\end{equation}
Most of time, the paraxial approximation helps to simplify the mathematics and obtain analytic expressions. In this case, Equations (\ref{eq:GeneralEq2}) and (\ref{eq:F2}) provides much clearer picture and its results are usually accurate enough to describe the light sheet. If more precision is required and the paraxial condition does not meet, one should work with Eq. (\ref{eq:GeneralEq}). 

\section{Diffracting Light Sheets}  \label{sec:DLS}

We start from description of diffracting light sheets, because understanding the physics of the diffraction effect is helpful to design "nondiffracting" light sheets with extended length. For numerical computation, one can follow Eqs. (\ref{eq:GeneralEq})-(\ref{eq:Nomalization}). In order to obtain some insightful analytic expressions, we here take 2D Gaussian beam as an example to illustrate the general relation between the light sheet parameters and the conclusions can be extended to general non-Gaussian sheets. Surprisingly, I have not found any literature that provides a complete analytic expression for 2D Gaussian beam propagation, which is different from the scanning Gaussian-beam light sheet. 

For a 2D Gaussian light sheet with a thickness $w_0$ [full width at half maximum (FWHM): $1.18 w_0\simeq w_0$] at waist, its angular spectrum is given by
\begin{equation}
G(k_x)=\sqrt{\frac{w_0}{\sqrt{2\pi}}}\exp\big[\frac{-w_0^2k_x^2}{4}\big],
\label{eq:GaussianSpectrum}
\end{equation}
which has a power spectrum ($|G(k_x)|^2$) width of about $2/w_0$ (FWHM: $2.35/w_0 \simeq 2/w_0 $). Under the slowly-varying envelope approximation of paraxial wave propagation, we have the z-dependant angular spectrum 
\begin{equation}
\begin{split}
\tilde{G}(k_x,z)=\sqrt{\frac{w_0}{\sqrt{2\pi}}}\exp\big[-(\frac{w_0^2}{4}+i\frac{z}{2k_0})k_x^2\big] e^{ik_0z}.
\end{split}
\label{eq:GaussianF2}
\end{equation}
Assuming the $k_x$ spectrum is confined within $[-k_0,k_0]$ (\textit{i.e.}, $2/w_0 < 2k_0$),  we obtain the 2D Gaussian beam mode from Eq.~(\ref{eq:GeneralEq2})
\begin{equation}
\begin{split}
g(x,z)=\sqrt{\frac{2}{\sqrt{2\pi}w(z)}}\exp\big[\frac{-x^2}{w^2(z)}\big]\exp\big[ik_0z\big] \\ 
\times\exp\big[i\frac{k_0x^2}{2R(z)}\big]\exp\big[-i\arctan(\frac{z}{2z_0})\big],
\end{split}
\label{eq:1DGaussianBeam}
\end{equation}
where $z_0=\pi w_0^2/\lambda$ is the Rayleigh range, $w(z)=w_0\sqrt{1+(z/z_0)^2}$ is the beam thickness  at distance z, and $R=z[1+(z/z_0)^2]$ is the radius of curve of the wavefront. We note that the 2D Gaussian beam in Eq. (\ref{eq:1DGaussianBeam}) is different from the well-known formalism of 3D Gaussian beam where the amplitude is proportional to $1/w(z)$ \cite{LaserElectronics}. The light sheet intensity distribution is give by
\begin{equation}
I_g(x,z)=|g(x,z)|^2=\frac{2}{\sqrt{2\pi}w(z)}\exp\big[\frac{-2x^2}{w^2(z)}\big].
\label{eq:1DGaussianIntensity}
\end{equation}
At $z=\sqrt{3}z_0$ the beam thickness increases by a factor of 2 and the central intensity drops by $1/2$. Then the Gaussian light sheet propagation length is  
\begin{equation}
L_0=2\sqrt{3}z_0=2\sqrt{3}\pi w_0^2/\lambda, 
\label{eq:1DGaussianBeamLength}
\end{equation}
which scales as $w_0^2$. At the diffraction limit $w_0=\lambda/2$, the propagation distance becomes only $L_0=\sqrt{3}\pi\lambda/2\simeq 2.72 \lambda$. As this view is too small for single cells, Gaussian-beam light sheets is not suitable for sub-cellar high resolution imaging. Shown in Eq. (\ref{eq:1DGaussianBeamLength}), as the thickness $w_0$ increases, the length $L_0\propto w_0^2$ increases significantly. Therefore, the practical resolution of a Gaussian light sheet is $>\lambda$. Equation (\ref{eq:1DGaussianBeamLength}) also suggests that the propagation length of a 2D Gaussian light sheet is improved by a factor of $\sqrt{3}$ as compared to that of a scanning Gaussian-beam light sheet with the same thickness. A spread light sheet is less diffractive then a scanning beam sheet. 

For other diffracting light sheets, one can compute the propagation wave following Eq. (\ref{eq:GeneralEq}). We numerically show that, for a square angular spectrum distribution with $k_x-$spectrum width $BW=\sqrt{8\pi}/w_0$, its light-sheet property is similar to that of a Gaussian light sheet with thickness $w_0$. Extending to a more general case, its full-width-at-half-maximum (FWHM) sheet thickness $w_0$ and the FWHM spectrum bandwidth $BW$ follows $BW w_0\simeq(2\sim5)$.  

\section{Light Sheets with Extended Length} \label{sec:NDLS}

\begin{figure*}
\centering
\fbox{\includegraphics[width=0.9\linewidth]{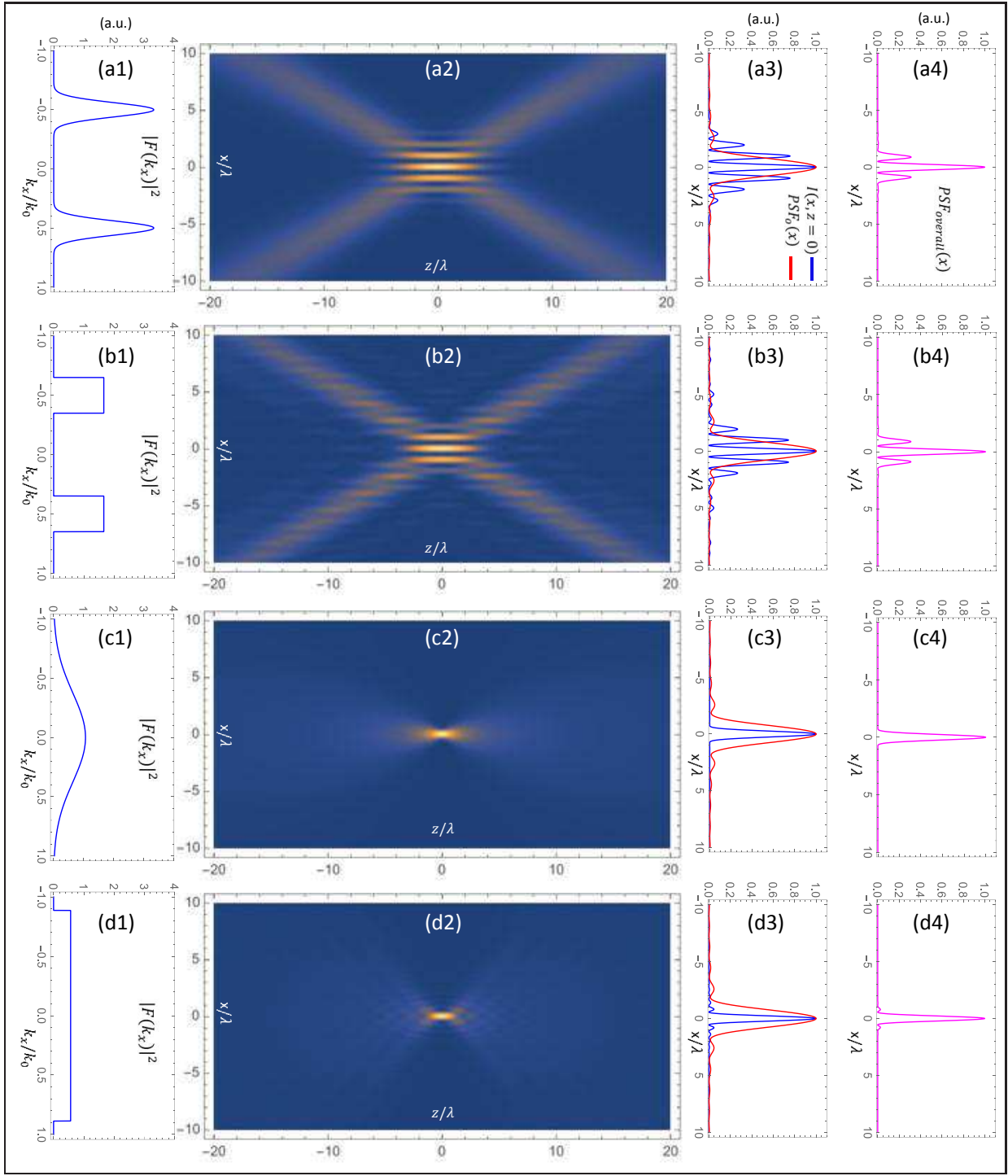}}
\caption{Light Sheets with FWHM thickness approaching the diffraction limit $d=0.50\lambda$. (a) Light sheet with two Gaussian spectrum bands: $w_0=2.65 \lambda$, $k_a=0.5 k_0$, FWHM sheet length $L=5.40\lambda$. (b) Light sheet with two rectangular spectrum bands: $BW=0.3 k_0$, $k_a=0.5 k_0$, FWHM sheet length $L=5.04\lambda$.  (c) Diffracting Gaussian light sheet: $w_0=0.42 \lambda$, FWHM sheet length $L=2.07\lambda$. (d) Diffracting light sheet with rectangular scpetrum: $BW=1.77 k_0$, FWHM sheet length $L=1.76\lambda$. (a1)-(d1): $k_x$ space power spectrum. (a2)-(d2): Light sheet intensity plots $I(x,z)$. (a3)-(d3): Light sheet intensity profiles $I(x,z=0)$ and the objective axial PSF [$PSF_o(x)$] with NA=0.8. (a4)-(d4): Overall PSF [$PSF_{overall}(x)$].}
\label{fig:1}
\end{figure*}

Now we turn to production of light sheets with extended length. To ensure a constant propagation $z$ number $\sqrt{k_0^2-k_x^2}$ in Eq. (\ref{eq:GeneralEq}), the angular spectrum takes $F(k_x)=F_+\delta(k_x-k_a)+F_-\delta(k_x+k_a)$. That is, an ideal diffraction-free (or non-diffracting) light sheet is a superposition of two 2D plane waves with $k_x=\pm k_a$. In a real situation, the $\delta$ function has to be replaced by a spectrum distribution with a finite bandwidth, which limits the diffraction-free propagation range. Therefore, a real ``non-diffracting" light sheet can be described as an interference of double slits in $k_x$-spectrum. Here the word ``non-diffracting" refers to a propagation length exceeding that of the diffracting light sheet with the same sheet thickness.   

We consider here the angular spectrum
\begin{equation}
\begin{split}
F_{s}(k_x)=\frac{1}{\sqrt{2}}[F_+(k_x)+F_-(k_x)] \\
=\frac{1}{\sqrt{2}}[F_0(k_x-k_a)+F_0(k_x+k_a)],
\end{split}
\label{eq:DoubleGaussianSpectrum1}
\end{equation}
where $F_0(k_x)$ is the spectrum centred at $k_x=0$. Taking $\beta=k_x\mp k_a$ and the paraxial approximation, we have
\begin{equation}
\begin{split}
\tilde{F}_{\pm}(k_x,z)e^{ik_xx}=F_0(\beta)e^{i(\beta\pm k_a)x}e^{i\sqrt{k_0^2-(\beta\pm k_a)^2}z} \\
\simeq e^{\pm i k_a x}e^{-i\frac{k_a^2}{2k_0}z} \tilde{F}_{0}(\beta,z)e^{\mp i\frac{k_a \beta}{k_0}z} e^{i\beta x},
\end{split}
\label{eq:DoubleGaussianSpectrum2}
\end{equation}
which is equivalent to 
\begin{equation}
\tilde{F}_{\pm}(\beta,z)=e^{\pm i k_a x}e^{-i\frac{k_a^2}{2k_0}z} \tilde{F}_{0}(\beta,z)e^{\mp i\frac{k_a \beta}{k_0}z}.
\label{eq:DoubleGaussianSpectrum3}
\end{equation}
Here $\tilde{F}_{0}(\beta,z)=F_0(\beta) e^{ik_0z} e^{-i\frac{\beta^2}{2k_0}z}$ follows the definition in Eq. (\ref{eq:F2}). Then from Eq. (\ref{eq:GeneralEq2}) we get
\begin{equation}
f_{\pm}(x,z)=e^{\pm i k_a x}e^{-i\frac{k_a^2}{2k_0}z} f_0(x\mp\frac{k_a}{k_0}z,z).
\label{eq:DoubleGaussianSpectrum3}
\end{equation}
Finally the non-diffracting light sheet is obtained as
\begin{equation}
\begin{split}
f_s(x,z)=\frac{1}{\sqrt{2}}[f_+(x,z)+f_-(x,z)] \\
= \frac{1}{\sqrt{2}} e^{-i\frac{k_a^2}{2k_0}z} \{f_0(x-\frac{k_a}{k_0}z,z)e^{i k_a x}+f_0(x+\frac{k_a}{k_0}z,z)e^{-i k_a x}\}.
\end{split}
\label{eq:NDGLS}
\end{equation}
The light sheet intensity distribution is given by
\begin{equation}
I_s(x,z)= \frac{1}{2}  |f_0(x-\frac{k_a}{k_0}z,z)e^{i k_a x}+f_0(x+\frac{k_a}{k_0}z,z)e^{-i k_a x}|^2.
\label{eq:NDGLSI}
\end{equation}
Equations (\ref{eq:NDGLS}) and (\ref{eq:NDGLSI}) show there is interference between the two diffracting sheets $f_\pm(x,z)$. How could the interference between two diffracting light sheets give non-diffracting feature? We can understand this in the following. We know for a single diffracting light sheet described in Sec. \ref{sec:DLS}, the sheet length or the propagation length is shorter than three wavelength as we make the sheet thickness close to the diffraction limit $\lambda/2$. Now let's design our non-diffracting light sheet to beat this diffraction effect with two interfering sheets. Each sheet is much thicker than the diffraction limit such that their propagation length is much longer than their wavelength. The interference modulates the intensity distribution and concentrates more energy on the central peak (with a factor of 2 enhancement as compared to a single beam). If we design the sheet properly so that most energy is in the central peak whose thickness is determined by 
\begin{equation}
d_s=\frac{\pi}{2 k_a},
\label{eq:NDLSThickness}
\end{equation}
which can be easily made shorter than the diffraction limit $\lambda/2$. In this way we create a light sheet with a thickness below the diffraction limit and a long propagation length that is beyond a single diffracting light sheet.  

To illustrate the idea, in the following we take double-Gaussian bands as an example and derive some analytic expressions. We take here the angular spectrum
\begin{equation}
\begin{split}
G_s(k_x))=\frac{1}{\sqrt{2}}[G_+(k_x)+G_-(k_x)] \\
=\frac{1}{\sqrt{2}}\{\sqrt{\frac{w_0}{\sqrt{2\pi}}}\exp\big[\frac{-w_0^2(k_x-k_a)^2}{4}\big]   \\
+\sqrt{\frac{w_0}{\sqrt{2\pi}}}\exp\big[\frac{-w_0^2(k_x+k_a)^2}{4}\big]\},
\end{split}
\label{eq:DoubleGaussianSpectrum1}
\end{equation}
where each band has a spectrum width $2/w_0$, corresponding to a sheet thickness $w_0$ of a diffracting Gaussian light sheet. To ensure the double-Gaussian spectrum is confined within $[-k_0,k_0]$, we require $k_a+1/w_0<k_0$. Following the previous discussion we have the intensity distribution 
\begin{widetext}
\begin{equation}
\begin{split}
&I_{gs}(x,z)= \frac{1}{2}  |g(x-\frac{k_a}{k_0}z,z)e^{i k_a x}+g(x+\frac{k_a}{k_0}z,z)e^{-i k_a x}|^2 \\
&=\frac{1}{\sqrt{2\pi}w(z)} \big | \exp\big[\frac{-(x-\frac{k_a}{k_0}z)^2}{w^2(z)}\big] \exp\big[i\frac{k_0(x-\frac{k_a}{k_0}z)^2}{2R(z)}\big] \exp[ik_ax]
+\exp\big[\frac{-(x+\frac{k_a}{k_0}z)^2}{w^2(z)}\big] \exp\big[i\frac{k_0(x+\frac{k_a}{k_0}z)^2}{2R(z)}\big]\exp[-ik_ax] \big |^2
\label{eq:NDGaussianSheet}
\end{split}
\end{equation}
\end{widetext}
The first term on the right side describes a 2D gaussian beam with a propagation vector \{$k_a, \sqrt{k_0^2-k_a^2}$\}, \textit{i.e.}, with an angle $\theta=\arcsin (k_a/k_0)$ to the z axis. The second term describes a 2D Gaussian beam with a propagation vector \{$k_a, -\sqrt{k_0^2-k_a^2}$\}, \textit{i.e.}, with an angle $-\theta$ to the z axis. As discussed previously, the central peak has a thickness $d=\pi/(2 k_a) < w_0$. The diffraction-free propagation length can be determined from the overlap of the two beams
\begin{equation}
L_{s}=\min\{\frac{w_0}{\sin \theta}, L_0\cos \theta\}=\min\{\frac{k_0}{k_a} w_0, \sqrt{3(k_0^2-k_a^2)} w_0^2\}.
\label{eq:NDLSLength}
\end{equation}

\begin{figure*}
\centering
\fbox{\includegraphics[width=0.9\linewidth]{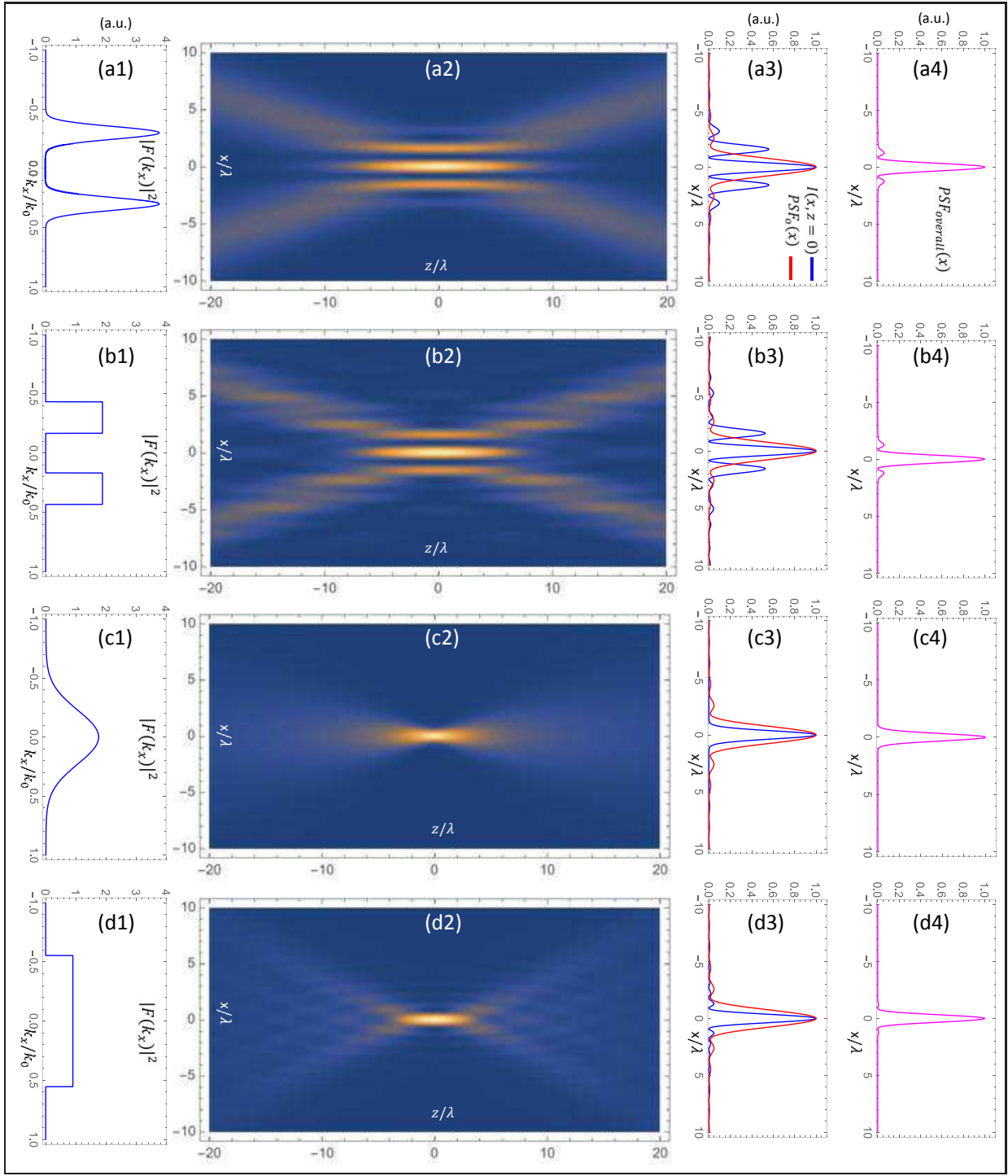}}
\caption{Light Sheets with FWHM thickness $d=0.84\lambda$. (a) Light sheet with two Gaussian spectrum bands: $w_0=3 \lambda$, $k_a=0.3 k_0$, FWHM sheet length $L=11.28\lambda$. (b) Light sheet with two rectangular spectrum bands: $BW=0.27 k_0$, $k_a=0.3 k_0$, FWHM sheet length $L=10.52\lambda$.  (c) Diffracting Gaussian light sheet: $w_0=0.70 \lambda$, FWHM sheet length $L=5.17\lambda$. (d) Diffracting light sheet with rectangular scpetrum: $BW=1.11 k_0$, FWHM sheet length $L=5.28\lambda$. (a1)-(d1): $k_x$ space power spectrum. (a2)-(d2): Light sheet intensity plots $I(x,z)$. (a3)-(d3): Light sheet intensity profiles $I(x,z=0)$ and the objective axial PSF [$PSF_o(x)$] with NA=0.8. (a4)-(d4): Overall PSF [$PSF_{overall}(x)$].}
\label{fig:2}
\end{figure*}

\begin{figure*}
\centering
\fbox{\includegraphics[width=0.9\linewidth]{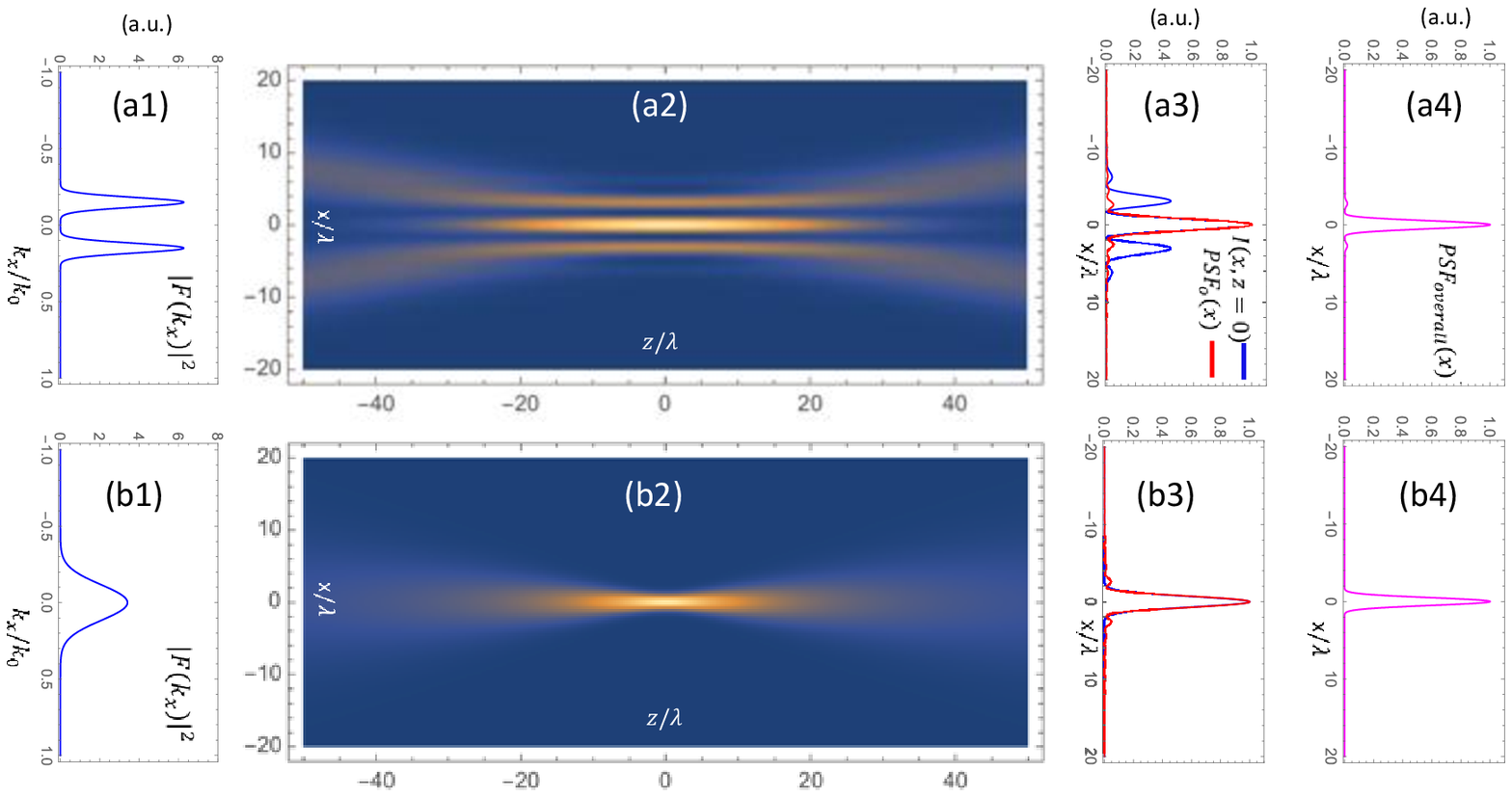}}
\caption{Light Sheets with FWHM thickness $d=1.60\lambda$. (a) Light sheet with two Gaussian spectrum bands: $w_0=5 \lambda$, $k_a=0.15 k_0$, FWHM sheet length $L=39.00\lambda$. (b) Diffracting Gaussian light sheet: $w_0=1.36 \lambda$, FWHM sheet length $L=19.90\lambda$. (a1)-(b1): $k_x$ space power spectrum. (a2)-(b2): Light sheet intensity plots $I(x,z)$. (a3)-(b3): Light sheet intensity profiles $I(x,z=0)$ and the objective axial PSF [$PSF_o(x)$] with NA=0.8. (a4)-(b4): Overall PSF [$PSF_{overall}(x)$].}
\label{fig:3}
\end{figure*}

For a single 2D Gaussian beam, as shown in Eq. (\ref{eq:1DGaussianBeamLength}),  its sheet thickness $w_0$ and propagation length $L_0$ is strong coupled ($L_0\propto w_0^2$) and cannot be independently tuned. For a non-diffracting light sheet created by interfering two 2D Gaussian beams, its sheet thickness $d_s$ and propagation length $L_s$ are jointly determined by two parameters $k_a$ and $w_0$ as shown in Eqs. (\ref{eq:NDLSThickness}) and (\ref{eq:NDLSLength}), providing more degrees of freedoms as compared to the single Gaussian-beam diffracting light sheets.

There are tradeoffs for optimizing a light sheet performance. Sometimes, we prefer a long light sheet that requires narrowband spectrum bands, but the narrower bands result in more side lobes and higher phototoxicity. Wider bands reduce phototoxicity but lead to a shorter sheet length. The number of side lobes also depends on the separation ($2k_a$) of the two bands. Equations (\ref{eq:NDLSThickness}) and (\ref{eq:NDLSLength}) are for designing non-diffracting light sheets with two Gaussian bands, and can also provide initial design parameters for other non-Gaussian-bands light sheets. The optimization can be done by finely tuning the two parameters, the bandwidth and separation, following the integral of Eq.  (\ref{eq:GeneralEq}). For achieving a thiner light sheet with longer length comparing to a diffracting single-band sheet, we can refer to the following two inequality $d_s > w_0$ and $L_s > L_0$. Equations (\ref{eq:NDLSThickness}) and (\ref{eq:NDLSLength}) can be extended to double rectangular bands with $BW=\sqrt{8\pi}/w_0$.

For a light-sheet microscope with the objective lens placed along the $x$ axis, the overall three-dimensional (3D) point spread function (PSF) is determined by
\begin{equation}
\mathrm{PSF}_{overall}(x, y, z)=\mathrm{PSF}_o(x,y,z) |f_s(x,z)|^2,
\label{eq:PSFoverall}
\end{equation}
where $\mathrm{PSF}_o(x,y,z)$, the PFS of the objective, is given by 
 \begin{equation}
\begin{split}
\mathrm{PSF}_{o}(x, y, z)=\big|\frac{1}{2\pi \sqrt{\pi}k_0} \\
\times \int \int_{k_z^2+k_y^2\leq NA^2 k_0^2}e^{i(k_zz+k_yy)}e^{i\sqrt{k_0^2-k_z^2-k_y^2}x}dk_zdk_y \big|^2,
\end{split}
\label{eq:PSFobjective}
\end{equation}
 where $NA$ is the numerical aperture of the objective.

Figure \ref{fig:1} shows numerical results of light sheets with FWHM thickness approaching the diffraction limit $d=0.50\lambda$. (a) and (b) are light sheets with two Gaussian spectrum bands and two rectangular spectrum bands, respectively. For comparison, we also give the diffracting Gaussian light sheet and rectangular-spectrum light sheet with the same sheet thickness in (c) and (d), respectively. Figure \ref{fig:1}  confirms that a Gaussian spectrum and a rectangular spectrum give nearly identical light sheet performance, following the relation $BW=\sqrt{8\pi}/w_0$. In this case, the diffracting light sheets have the propagation length of only about $2\lambda$, while in the non-diffracting sheets it is extended to more than $5\lambda$. As compared to a Bessel beam whose energy is evenly distributed in its side lobes, in the non-diffracting light sheets there are only a few of confined side lobes [Fig. \ref{fig:1}(a3)]. Using an objective lens with NA=0.8 whose axial FWHM of PSF is $1.60 \lambda$, we achieve an overall PSF with axial FWHM $0.47 \lambda$, as shown in Fig. \ref{fig:1}(a4). This suggests a light sheet microscope has a much higher axial imaging resolution than a conventional microscope. The effect of the residual side peaks in the overall PSF can be removed with deconvolution method.

In Fig. \ref{fig:1}, we optimize the axial resolution and there are four visible side lobes. To reduce the number of side lopes for less photototoxicity, there are two choices: (1) to reduce $w_0$ or increase the spectrum bandwidth of each band, and (2) to reduce the band separation $k_a$. The first choice is not preferable because it also reduces the sheet propagation length. The second choice increases the sheet propagation length but with the cost of increasing the thickness. Figure \ref{fig:2} shows light sheets with less photototocity and fewer side lobes, by reducing $k_a$ and slightly increasing $w_0$. The light sheet thickness now becomes $d=0.84\lambda$. The non-diffracting light sheet length is extended to $L=11\lambda$. The side lobes in the overall PSF nearly disappear and its axial resolution is $0.72\lambda$. 

We can also design the light sheet with thickness matching the PSF of the objective, as shown in Fig. \ref{fig:3}(a), with further increasing $w_0$ to $5.0\lambda$ and reducing $k_a$ to  $0.15 k_0$. The non-diffracting light sheet now has a FWHM thickness of  $1.6\lambda$ and propagation length of $39\lambda$. The axial resolution of the overall PSF is $1.16\lambda$. The propagation length of a diffracting Gaussian light sheet with the same thickness is only $20\lambda$, as shown in Fig. \ref{fig:3}(b).

For practical implementation, we can generate light sheets with extended length by shining a plane or line wave beam on a double-slit mask which is placed on the back focal plane of a Fourier lens. In real situation, the finite aperture size of the lens should be taken into account \cite{FourierOptics}.

\section{Summary} \label{sec:Summary}

In summary, we have provided formalism for designing 2D light sheets with two symmetric bands in the $k_x$ space. For Gaussian-shape band structure, we provide analytic expressions under the paraxial approximation, which can be extended to rectangular spectrum with $BW=\sqrt{8\pi}/w_0$. As compared to the diffracting light sheet with the same thickness, the sheet propagation length is extended significantly. In the diffracting light sheet, the sheet propagation length is proportional to square of its thickness ($L_0=2\sqrt{3}\pi w_0^2/\lambda$) and these two numbers are strongly coupled. In the non-diffracting case, the light sheet thickness and propagation length are determined by the spectrum bandwidth ($BW$) and their separation ($k_a$), which provide much flexibility for optimizing the sheet parameters. The results are confirmed by numerical simulations. As compared to scanning Bessel-beam light sheets and dithering lattice light sheets, the spatially spreading light sheets described here is simpler and less phototoxic. LSM with these light sheets will have important applications in 4D live-cell imaging with high spatial-temporal resolution.  
\\

\section*{Acknowledgments}

The work was supported by Light Innovation Technology Ltd through a research contract with the Hong Kong University of Science and Technology (HKUST) R and D Corporation Ltd (Project code 16171510), and the Offices of the Provost, VPRD and Dean of Science, HKUST (project no. VPRGO12SC02).

\end{document}